\documentstyle[12pt,epsf]{article}

\newcommand{\be}{\begin{equation}}
\newcommand{\ee}{\end{equation}}

\def\aprge{\buildrel > \over {_{\sim}}}
\begin{document}
\topmargin 0pt
\oddsidemargin=-0.4truecm
\evensidemargin=-0.4truecm
\renewcommand{\thefootnote}{\fnsymbol{footnote}}
\newpage
\setcounter{page}{1}
\begin{titlepage}     
\vspace*{-2.0cm}
\begin{flushright}
\vspace*{-0.2cm}
hep-ph/9902312\\
IC/99/9\\
\end{flushright}
\vspace*{-0.5cm}

\begin{center}
{\Large \bf Testing the Solar Neutrino Conversion with Atmospheric
Neutrinos}
\vspace{0.5cm}

{O. L. G. Peres$^{1}$ and  A. Yu. Smirnov$^{1,2}$\\
{\em (1) The Abdus Salam International Centre for Theoretical Physics,  
I-34100 Trieste, Italy }\\
{\em (2) Institute for Nuclear Research of Russian Academy 
of Sciences, Moscow 117312, Russia}

}
\end{center}

\vglue 0.8truecm
\begin{abstract}

Neutrino oscillations with  parameters $\Delta m^2_{\odot} =  (2 - 20)
\cdot 10^{-5}$ eV$^2$, $\sin^2 2\theta_{\odot} > 0.65$,
relevant for large mixing MSW solution of the solar neutrino problem 
can  lead to an observable (up to 10 - 12 \%) excess of the 
e-like events in the sub-GeV atmospheric neutrino sample. 
The excess has a weak zenith angle dependence 
in the low energy part of the sample and strong zenith 
angle dependence in the high energy part. 
The excess rapidly decreases with energy of neutrinos, it 
is suppressed in the multi-GeV sample. These 
signatures allow one to disentangle the effect of the oscillations 
due to solar $\Delta m^2$  from other possible explanations of the excess.  
The up-down  asymmetry of the excess may change the sign with energy
being positive in the sub-GeV region and negative in the multi-GeV range. 
Predicted properties of the excess are  in agreement with the 
SuperKamiokande data.

\end{abstract}
\end{titlepage}
\renewcommand{\thefootnote}{\arabic{footnote}}
\setcounter{footnote}{0}
\newpage
\section{Introduction}

The SuperKamiokande (SK) collaboration~\cite{SuperK,skwin99} 
continues to accumulate the data which strengthen 
the evidence of the muon neutrino oscillations. 
Whole variety of the  data can be fitted well assuming 
$\nu_\mu\leftrightarrow \nu_\tau$ channel with the maximal or
close to maximal mixing:
\begin{equation}
\Delta m^2_{atm} = (1.5 - 8) \times 10^{-3} {\rm eV}^2~, ~~~
\sin^2 2\theta_{atm} > 0.8. 
\label{atmdata}
\end{equation}
At the same time, there are some facts which indicate that
oscillations of the atmospheric neutrino is  not reduced 
completely to two neutrino case  and the electron neutrino is
involved  in the oscillations too.  The $\nu_{\mu} 
\leftrightarrow \nu_e$ oscillations as the sub-leading mode are   
possible and probably desired. 

\begin{itemize}

\item

The data shows an  excess of the e-like events 
in the sub-GeV sample. 
The SK collaboration accounts the excess 
by up-scaling
the overall normalization of the neutrino fluxes.  
However, recent cosmic ray measurements~\cite{BESS} 
indicate that possible increase  of the
neutrino fluxes was probably overestimated. Moreover, it seems that
the excess depends on the neutrino energy (which is impossible to
explain by overall normalization). The largest excess is  in the  low 
energy part of the sample ($p < 0.4$~GeV) and it is smaller in the 
high energy part.

\item 

There are some  indications of the up-down asymmetry 
of the e-like events which  changes with energy.   
The asymmetry is positive in 
the low momenta region 
and  it tends to be  negative at high energies (multi-GeV range). 

\item 

The data show strong zenith angle dependence of the 
$\mu$-like events which imply relatively small oscillation effect for
down-going neutrinos.  In the  two neutrino
framework   it is difficult 
to reconcile this fact with rather low value of the 
double ratio $R_{\mu/e} \equiv  (\mu/e)/(\mu/e)_{MC}$ 
integrated over zenith angle~\cite{LoSecco}.    
(Although recent data give higher $R_{\mu/e}$
the problem is not completely settled down.) 
In the case of three neutrino oscillations the
$\nu_{\mu}$-signal  can be further suppressed and the $\nu_{e}$-signal
enhanced.

\end{itemize}


There have been a number of studies of the atmospheric neutrino oscillations 
with the three (or more)  neutrino mixing
\cite{threenu1,threenu2,threenu3,threenu4,four1,ADLS,yasuda}. Most of
them where performed in the framework of the so called ``one level
dominating scheme''  when the mass splitting between two lightest states,  
$\Delta m^2_{21}$, is neglected. In this case the oscillations in the
sub-leading
channel occur due to mixing of the electron neutrino in the heaviest 
mass state  with mass splitting $\Delta m^2_{atm}$. 
The effects of the sub-leading channel are restricted by the 
CHOOZ result~\cite{CHOOZ}. 
They are reduced to vacuum oscillations for the sub-GeV sample. 
For the multi-GeV sample the Earth matter effect becomes important
which may enhance  or suppress the oscillations~\cite{ADLS}.

In several papers all  mass splittings have been taken into account. 
In Ref.~\cite{yasuda} generic 3$\nu$ effects have been considered
with relatively large $\Delta m^2_{21}$ for lightest states (outside
the region of solar neutrino solutions).
In Ref.~\cite{thun} the attempts have been made to  explain the zenith angle
dependence of the $\mu$-like events by the $\nu_{\mu} \leftrightarrow
\nu_e$ oscillations  with small $\Delta m^2_{21} \sim 10^{-4}$ eV$^2$
and large mixing. The channel $\nu_{\mu} \leftrightarrow \nu_{\tau}$ with 
large $\Delta m^2_{32} \sim 0.2$ eV$^2$ implied by the LSND result 
leads to overall suppression of the signal. It was marked that
for $\Delta m^2_{21} < 10^{-4}$ eV$^2$  the matter of the Earth significantly 
suppresses the oscillation effect. In this
scheme one expects also strong zenith angle dependence of the e-like
events which contradicts the SK data.

It was marked in Ref.~\cite{Fogli} that the effect 
of the sub-leading oscillations driven by 
$\Delta m^2_{\odot}$ responsible for the solar neutrino deficit 
is significant only for the sub-GeV events and the size of effect 
at the level of the statistical errors. In Ref.~\cite{kim} it was
argued that the excess of the e-like events in the sub-GeV sample favors
large mixing MSW solution of the solar neutrino problem. However
analysis of 535 days of the SuperKamiokande data in
Ref.~\cite{sakai99} leads to conclusion that there is no difference
between the large and small mixing solutions.

The impact  of the solar neutrino sector on the 
atmospheric neutrino fluxes has been also studied in the 
context of exact parity model~\cite{brunn}. 



In this paper we will study in details the effects of 
oscillations of the atmospheric neutrinos with parameters 
($\Delta m^2_{\odot},\sin^2 2\theta_{\odot}$)
of the large mixing MSW solution of the solar neutrino
problem.
The analysis of recent  data (including the spectral and 
zenith angle information) leads to the following values~\cite{skwin99,bks} 
\be
\Delta m^2_{\odot} = (2 - 20)\cdot 10^{-5}~ {\rm eV}^2~,~~
\sin^2 2\theta_{\odot} =  0.65 - 0.95~.
\label{solar}
\ee
This region corresponds to two neutrino mixing. It is 
slightly modified if there is an
admixture of 
$\nu_e$ flavor in the third state which satisfies the 
CHOOZ bound~\cite{CHOOZ}: $|U_{e3}|^2 = \sin^2 \theta_{13} < 0.05$. 

There are two motivations of present  study:

(1).  Recent SK data give some indications in favor of the  
large mixing MSW solution (flat distortion of the recoil electron spectrum,  
day-night asymmetry,  flat zenith angle distribution of
the night signal, etc.)~\cite{skwin99,bks1} .

(2). Studies of the atmospheric neutrinos enter now new  stage  of 
precision measurements when the data become sensitive to sub-leading
effects.

The paper is organized as follows. 
In Sec. 2 we find general expressions for the 
atmospheric neutrino fluxes in presence of the 
three neutrino oscillations. In Sec. 3 we discuss properties of 
relevant two neutrino probabilities.  
In sect. 4 we calculate the excess of the e-like events
for different energy ranges and study its properties. 
In sect. 5 we consider influence of the sub-leading oscillations on
the $\mu-$like events and $R_{\mu/e}$. In Sec. 6 we discuss the results and 
perspectives.

\section{Three-flavor oscillations of atmospheric neutrinos}

We consider the three-flavor neutrino system  with  
hierarchical mass squared differences: 
$\Delta m^2_{21} = \Delta m^2_{\odot}<< \Delta m^2_{31} = \Delta
m^2_{atm}$ (see Eqs.~(\ref{atmdata},\ref{solar})). 
The evolution of the neutrino vector of state $\nu_f \equiv (\nu_e,
\nu_{\mu}, \nu_{\tau})^T$ is described by the equation 
\be
i \frac{d \nu_f}{dt} =
\left( \frac{U M^2 U^\dagger}{2 E} + V \right) \nu_f, 
\label{evolution}
\ee 
where $E$ is the neutrino energy and  
$M^2 = diag(0, \Delta m_{21}^2, \Delta m_{31}^2)$
is the diagonal matrix of neutrino mass squared eigenvalues. 
$V = diag(V_e, 0 ,0)$ is the matrix of matter-induced neutrino potentials
with $V_e = \sqrt 2 G_F N_e$, $G_F$ and $N_e$ being the Fermi constant 
and the electron number density, respectively.   
The mixing matrix $U$ is defined through $\nu_f = U \nu_{m}$, where $\nu_{m}=
(\nu_1, \nu_2, \nu_3)^T$ is the vector of neutrino mass
eigenstates. It can  be parameterized as 
$
U = U_{23} U_{13} U_{12}.   
$
The matrix $U_{ij}= U_{ij}(\theta_{ij})$ performs the rotation  
in the $ij$- plane by the angle $\theta_{ij}$. 
We have neglected possible CP-violation effects in the lepton sector 
which are suppressed in the case of the mass
hierarchy.

Let us introduce new states 
$\tilde{\nu} = (\nu_e, \tilde{\nu}_{2}, \tilde{\nu}_{3})^T$  
obtained by performing the $U_{23}$ - rotation:
$\nu_f = U_{23} \tilde{\nu}$. The Hamiltonian $\tilde{H}$ that 
describes the evolution of the vector $\tilde{\nu}$   can be found from 
Eq.~(\ref{evolution}): 
$$
\tilde{H} =  \frac{1}{2E} U_{13} U_{12} M^2
U^\dagger_{12} U^\dagger_{13} ~ + ~ V ~.
$$ 

Let us assume that mixing of the electron neutrino in the heavy
state is negligible,\footnote{
We comment of the effect of this mixing in Section 6. Results of the 
detailed studies will be published elsewhere~\cite{RS}.}
so that $U_{13} \approx 1$. 
 In this case    
we get  explicitly 
\be
\tilde{H} \approx  \left(\begin{array}{ccc}
s_{12}^2 \Delta m_{21}^2/2E + V_e   &  s_{12}c_{12}\Delta m_{21}^2/2E 
& 0  \\
s_{12}c_{12} \Delta m_{21}^2/2E & c_{12}^2 \Delta m_{21}^2/2E  &   0  \\
0 &  0  & \Delta m_{31}^2/2E
\end{array}\right)\, , 
\label{matr1}
\ee
($c_{12} \equiv \cos \theta_{12}$, $s_{12} \equiv \sin \theta_{12}$, 
etc.)  
According to Eq.~(\ref{matr1}),  the $\tilde{\nu}_3$ state decouples from the
rest of the system
and evolves independently. Therefore the S-matrix (the matrix of  
amplitudes) in the basis $(\nu_e, \tilde{\nu}_2, \tilde{\nu}_3)$  
has the following form :
\be  
S \approx 
\left(\begin{array}{ccc} 
A_{ee}   & A_{e2}    & 0 \\
A_{2e}   & A_{22}    &   0    \\
0        & 0         & A_{33} 
\end{array}
\right) ~~ ,~~
\label{matr2}
\ee
where 
\be 
A_{33} = \exp(-i\phi_{3})\,, \quad \quad      
\phi_{3} = \frac{\Delta m_{31}^2 L}{2E},  
\label{phase}
\ee
and  $L$ is the total distance traveled by the neutrinos. 
The $(\nu_e, \tilde{\nu}_2)$ subsystem 
evolves according to the  2$\times$2 Hamiltonian 
($\nu_e - \tilde{\nu}_2$ sub-matrix in Eq.~(\ref{matr1})).  
The latter  depends on  the potential $V_e$, mixing
angle $\theta_{12}$ and 
the mass squared difference $\Delta m^2_{21}$. 
Let us denote by  
\be
P_2 \equiv |A_{e2}|^2 = |A_{2e}|^2 =    
1 - |A_{ee}|^2 = 1 - |A_{22}|^2
\ee 
the probability of the  $\nu_e \leftrightarrow
\tilde{\nu}_{2}$ oscillations.  For antineutrinos we have
$\overline{P}_2=P_2(-V_e)$.

The $S$-matrix in the flavor basis can be obtained from 
Eq.~(\ref{matr2}) by $U_{23}$-rotation: $U_{23}S U_{23}^{\dagger}$. 
It gives
the probabilities  of flavor oscillations
as 
$P(\nu_\alpha \to \nu_\beta)=|(U_{23} S U_{23}^{\dagger}) _{\alpha
\beta}|^2$. The appropriate probabilities equal  
\be
P(\nu_e\to \nu_e) = 1 - P_2 ~~, 
\label{ee}
\ee
\be
P(\nu_e\leftrightarrow \nu_\mu) = P(\nu_\mu \leftrightarrow \nu_e) = c_{23}^2  P_2 ~~, 
\ee
\be
P(\nu_\mu\to\nu_\mu ) = 1  - c_{23}^4 P_2 - 
2s_{23}^2 c_{23}^2\,\left[1 - \sqrt{1 - P_2} \cos \phi
\right]~~, 
\label{numu}
\ee
where 
$$
\phi \equiv \phi_3 - \phi_2  
$$
is the phase difference of the amplitudes $A_{22}$ and $A_{33}$:  
$\phi_2 \equiv arg{A_{22}}$ is the phase of $A_{22}$; 
$\phi_3$ is defined in Eq.~(\ref{phase}).

Using the probabilities given in Eqs.~(\ref{ee}-\ref{numu}) one can find
modifications of the atmospheric neutrino fluxes due to the oscillations. 
Let  $F_e^0$ and $F_{\mu}^0$ be the electron and muon neutrino fluxes at 
the detector in the absence of  oscillations. Then the fluxes in the
presence of  oscillations can be written as 
\be
F_e = F_e^0 \left[ 1 + P_2(r c_{23}^2 - 1)\right]~~, 
\label{fluxe}
\ee
\be 
F_{\mu} = F_{\mu}^0 \left[ 1 - \frac{c_{23}^2}{r} 
\left({r c_{23}^2 - 1} \right) P_2   
-  \frac{1}{2} \sin^2 2\theta_{23} 
\left( 1 - \sqrt{1 - P_2} \cos \phi  \right) \right], 
\label{fluxmu}
\ee
where 
$$
r(E, \Theta_\nu) = \frac{F_{\mu}^0(E, \Theta_\nu)}{ F_e^0(E, \Theta_\nu)}
$$ 
is the ratio of the original muon and electron neutrino fluxes. Here
$\Theta_\nu$ is the neutrino zenith angle.

For antineutrinos $P_2$ should be substituted by $\overline{P}_2$ 
in Eqs~(\ref{ee} -  \ref{fluxmu}).   
The solution of the solar neutrino problem implies that the 
resonance is in the neutrino channel,  therefore the mixing and 
the transition  probability are smaller for antineutrinos: 
$P_2 > \overline{P}_2$.

As follows from Eq.~(\ref{fluxe}) the effect of oscillations on the electron
neutrino flux 
is proportional to the factor $(r c_{23}^2 - 1)$. 
Therefore 
one can have either an excess or a deficiency of
the e-like events depending on  values of $r$ and
$c_{23}$. The ratio
$r$ depends both on the zenith angle and on the neutrino energy. 
For  $r = 2$ which corresponds to the 
sub-GeV sample, there will be an excess of e-like events for 
$\theta_{23} < 45^\circ$ and a deficiency for $\theta_{23} >  45^\circ$.
The SK best fit is  $\theta_{23} = 45^\circ$; in this case there would be 
no deviation from the prediction for $r = 2$. 
In the multi-GeV range $r$ is typically 3 -- 3.5 in the 
vertical direction with averaged over zenith angle value about 2.5.  

\section{The transition probability}

We have calculated the two neutrino transition probabilities $P_2$,
$\overline{P}_2$ using the parameterization of the  distribution of density
in the Earth from Ref.~\cite{earthmodel}. For the analysis of the
results, it is worthwhile to consider transitions of neutrinos in the
Earth as oscillations in medium which consists of  several layers with
constant densities.  This, in fact, gives rather good approximation
to the exact results. 

The depth  and the length of oscillations are determined by 
\be
\sin^2 2\theta_m =   \sin^2 2\theta_{12}  
\left(\frac{\Delta m^2_{21}} {2E \Delta H } \right)^2~,~~~~~~   
l_m = \frac{2\pi}{\Delta H},
\ee
where 
\be
\Delta H  =  
\sqrt{\left(\cos 2\theta_{12} \frac{\Delta m^2_{21}}{2E} -V_e\right)^2 
+ \left(\sin 2\theta_{12} \frac{\Delta m^2_{21}}{2E} \right)^2  }
\ee
is the level splitting (difference between  the eigenvalues 
of $H$). The resonance energy equals 
\be
E_R = \frac{\Delta m^2_{21} \cos 2\theta_{12}}{2V_e} = 0.17\;{\mbox GeV} 
\left( \frac{\Delta m^2_{21}}{5\cdot  10^{-5}{\mbox eV}^2 } \right)
\left( \frac{2.0g/cm^3}{Y_e\rho} \right)
\cos 2\theta_{12} .
\ee
In the mantle, for typical value 
$\Delta m^2_{21} = 5 \cdot 10^{-5}$~eV$^2$ 
and for $\sin^2 2\theta_{12}=0.8$ we get  $E_R~=~0.08$~GeV
which is below the threshold of the sub-GeV range. Therefore  
for  $\Delta m^2_{21} < 5 \cdot 10^{-5}$ eV$^2$ the oscillations are 
in the matter dominated regime when the potential is substantially
larger than the kinetic term: $V \gg \Delta m^2 / 2E$. 
In this case, the depth of oscillations is roughly proportional to 
$(\Delta m^2)^2$, and the oscillation length, $l_m$, is close to the refraction
length, $l_0$, and only weakly depends on the energy: 
\be
\sin^2 2\theta_m  \sim  \sin^2 2\theta_{12} 
\left(\frac{\Delta m^2_{21}} {2E V_e} \right)^2~,~~~~~~  
l_m \approx l_0 = \frac{2\pi}{V_e}. 
\label{mdominate}
\ee
For the multi-GeV range the approximation of Eq.~(\ref{mdominate})
works for $\Delta m^2$ as big as $10^{-4}$~eV$^2$. 

We find that the maximal neutrino oscillation effect in the mantle is achieved 
at $\cos \Theta_{\nu}~\sim -0.35$ and the effect is zero at $\cos
\Theta_{\nu} \sim - 0.6$. For $\cos \Theta_{\nu} <  - 0.84$ neutrinos
cross both the  mantle and the core of the Earth. The  interplay of
the oscillations in the mantle and in the core leads to some
enhancement of the transition  probability  in spite of larger density
of the core.  The oscillation effects in the antineutrino channel 
are  smaller by factor 2 - 3. 

The expressions of the Eq.~(\ref{mdominate}) are valid for  small
$\sin^2 2\theta_m$. With increase of  $\Delta m^2$, the increase of
$\sin^2 2\theta_m$, and consequently, the probabilities is slowing down. In
the neutrino channel   the depth approaches one  in the resonance. In
the antineutrino channel the mixing also increases  but it is always
below vacuum mixing.

\section{Excess of the  e-like  events}

In what follows we will calculate the dependences of the excess of
e-like events on the zenith angle of electron, $\Theta_e$. The general
expression for the number of e-like events, $N_{e}$ as a function of
$\Theta_e$ is

\begin{eqnarray}
  N_{e} \propto \sum_{\nu\overline{\nu}}
 \int  & \left. dE_\nu dE_e d(\cos\Theta_\nu) dh  
\;\;   F_e(E_{\nu},\Theta_{\nu}) \frac{d{ \sigma}}{dE_e} \right.  \\ 
 & \Psi(\Theta_e,\Theta_{\nu},E_{\nu})   \kappa_e(h,\cos\Theta_\nu,E_\nu)   
\varepsilon(E_e)  \;\;,
\label{event1}
\end{eqnarray}

\noindent where $F_e$ is the atmospheric $\nu_e$-flux
at the detector given in Eq.~(\ref{fluxe}) (the fluxes 
$F_e^0$ and $F_{\nu}^0$ without oscillations are taken from 
Ref.~\cite{flux});
$d\sigma / dE_{e}$ are the differential cross
sections taken from  Ref.~\cite{LLS},  $\kappa_e$ is the normalized
distribution of neutrino production points, $h$ is the height of
production, $\epsilon(E_e)$ is the detection efficiency of the
electron, $\Psi$ is the ``dispersion'' function which describes
deviation of lepton zenith angle from the neutrino 
zenith angle~(~For details see Ref.~\cite{compute}).

Notice that the integration over the neutrino zenith angle and
neutrino energy 
leads to a significant smearing of the 
$\Theta_{\nu}$ dependence.   
The average angle between the neutrino and the outgoing charged lepton  is
almost $60^{\circ}$ in the sub-GeV range  
and it is about 
$15^{\circ} - 20^{\circ}$ in the multi-GeV region. 
Neutrinos and antineutrinos of a given flavor are not distinguished 
in the atmospheric neutrino experiments, 
so that  the neutrino and antineutrino signals are summed in 
Eq.~(\ref{event1}) which also leads to weakening of the oscillation effect.

According to Eq.~(\ref{fluxmu}) and Eq.~(\ref{event1}) the relative
excess of the e-like events, $\epsilon_e$,  can be represented as  
\be 
\epsilon_e \equiv \frac {N_e}{N_e^0} - 1 \approx 
\overline{P} (\Theta_e) (\overline{r}(\Theta_e) c_{23}^2 - 1), 
\label{eps}
\ee  
where $\overline{P} (\Theta_e)$ is the probability  averaged over appropriate
energy 
and zenith angle intervals as well as over 
neutrinos and antineutrinos; $\overline{r}(\Theta_e)$ is the 
effective ratio of the electron and muon neutrino fluxes 
for a given energy and angle intervals.

The up-down asymmetry is given by 
\be
A^{U/D}_{e}  =  \frac
{N_{e}^{U} - N_{e}^{D}}
{N_{e}^{U} + N_{e}^{D}}~, 
\ee
where 
\be
N_{e}^{U} = \int_{- 1.0}^{-0.2} d \cos \Theta_e
N_{e}(\Theta_e)~,~~~ 
N_{e}^{D} = \int_{0.2}^{1.0} d \cos\Theta_e N_{e}(\Theta_e)~, 
\ee
and $N_{e}(\Theta_e)$ are  given  in Eq.~(\ref{event1}). 
This definition includes both asymmetry 
of the original neutrino flux (in particular, due to 
geomagnetic effect) and the asymmetry due to oscillations. The
asymmetry due to oscillations only can be estimated as 

\be 
A^{U/D}_{osc} \approx \frac{\epsilon_e^{up} - \epsilon_e^{down}}{2} ,
\ee
where $\epsilon_e^{U}$ and $\epsilon_e^{D}$ are values of the excess 
integrated over same zenith angle bins as in Eq.~(21). 

Let us consider the sub-GeV events which correspond 
to the limit of integration in Eq.~(\ref{event1}), $p < 1.33$~GeV. 
In Fig.~1  we show the zenith 
angle dependences of the excess of the e-like  events for 
$\sin^2 2\theta_{12} = 0.9$,  $\sin^2 \theta_{23} = 0.8$  and  
different values of $\Delta m^2_{21}$.  The following remarks are in
order.

(i). The excess  increases rapidly with 
$\Delta m^2_{21}$ in correspondence with behaviour of the 
probability. In the first  vertical upward bin the excess 
can reach $12.3$ \% for $\Delta m^2_{21} = 2\cdot  10^{-5}$
eV$^2$. The excess integrated over all bins  for different
values of $\Delta m^2_{21}$ is  shown in the Table~1.

\begin{table}
\begin{center}
\begin{tabular}{|c|c|c|}
\hline
$\Delta m^2_{21}$(eV$^2$)  &$\bar{\epsilon}_e (0.8)$ (\%) 
                         & $\bar{\epsilon}_e (0.8)$ (\%)   \\
\hline
                   & $\sin^2 2\theta_{12}=0.65$ & $\sin^2 2\theta_{12}=0.90$\\ 
\hline
$1. 10^{-5}$    &   0.1  &  0.2  \\
\hline
$4. 10^{-5}$    &   1.8  &  2.0  \\
\hline
$8. 10^{-5}$    &   4.7  &  4.8   \\
\hline
$2. 10^{-4}$    &   8.6  &  9.3  \\
\hline

\end{tabular}
\end{center}
\caption{The excess integrated over $\Theta_e$, $\bar{\epsilon}_e$, for
different choices of $\Delta m^2_{21}$ and
$\sin^2 2\theta_{12}$.}
\end{table}

(ii). The excess has rather weak zenith angle dependence.
For instance, the up-down asymmetry equals $A^{U/D}_{osc} = 1.5-4.0$ \% 
for $\Delta m^2_{21} \aprge 10^{-4}$ eV${^2}$, $\sin^2 2\theta_{12} = 0.9$
and $\sin^2 \theta_{23}~=~0.8$. 

(ii). The excess depends very weakly  on the solar mixing angle 
$\sin^2 2\theta_{12}$ (see Fig.~2 and Table 1). This is related to
weak dependence of  $\sin^2 2\theta_m$ on $\sin^2 2\theta_{12}$ in the
region of  parameters under consideration. The decrease of $\sin^2
2\theta_{12}$  is compensated by the shift of the resonance to larger
energies. 

(iv). The excess increases with the decrease of $\sin^2 2\theta_{23}$.  
According to Eq.~(\ref{eps}) the dependence of the excess on the
mixing angle of leading channel, $\theta_{23}$ is determined by the factor 
$ (\overline{r}(\Theta_e) c_{23}^2 - 1)$. Therefore for arbitrary value
of $\theta_{23}$ the excess can be found 
using $\epsilon_e (0.8)$ (the calculated excess for  
$\sin^2 2\theta_{23} = 0.8$; see Table 1):  
\be
\epsilon_e (\sin^2 2\theta_{23}) = 
\epsilon_e (0.8) \cdot  
\frac{c_{23}^2 \overline{r}(\Theta_e) - 1}{0.723 \overline{r}(\Theta_e) - 1}
\approx \epsilon_e (0.8) \cdot \frac{\cos 2 \theta_{23}}{0.45} .
\ee
For $\sin^2 2\theta_{23} = 0.7,\; 0.9,\;  0.95$ and 
$\overline{r} = 2$ the excess (in the units of $\epsilon_e (0.8)$)  
equals  1.23, 0.716, 0.49
correspondingly. The excess disappears when the mixing in the leading
channel approaches the maximal one. Notice, however that even for 
$\sin^2 2\theta_{23}=0.95$ an appreciable excess still survives.

The excess decreases with increase of  energy 
of the selected events. This corresponds to 
decrease of the mixing parameter in medium: $\sin^2 2\theta_m \propto  
1/E_\nu^2$ in the matter dominated range. 
In Fig.~3 and Fig.~4 we show the excess of the e-like events 
(as the function of the zenith angle $\Theta_e$) 
for the low energy part of the sub-GeV sample with  the electron
momentum $p < 0.4$ GeV, 
and for the high energy part of the sample with $p >  0.4$~GeV. 

For the low energy sample (Fig.~3) the excess  can reach  12 \%
in the upward vertical bin   
for $\Delta m^2_{21} = 2\cdot 10^{-4}$ eV$^2$. The excess has  weak
zenith angle dependence, e.g. for $\Delta m^2_{21}= 10^{-4}$~eV$^2$
it decreases from 9\% to 7.6 \% with increases of the zenith angle . 

The integrated excess
$\bar{\epsilon}_e$ equals $ 11.3$ \% and $8.4$ \%  for the two indicated
values of $\Delta m^2_{21}$.  

In the high energy sample (Fig.~4) the excess in the vertical 
bin also reaches 12 \% ($\Delta m^2_{21} = 2\; 10^{-4}$ eV$^2$),   
but it decreases rapidly with $\Delta m^2_{21}$. 
In this sample the zenith angle dependence is very strong, 
e.g. for $\Delta m^2_{21} = 10^{-4}$ eV$^2$ the excess decreases from 
5\% in the vertical upward bin to 0.7\% in the vertical 
down going bin, the asymmetry $A^{U/D}_{osc}=2-5 \% $ for 
 $\Delta m^2_{21} \aprge 10^{-4}$ eV$^2$.   
The integrated  excess is smaller than in the low energy sample:
$\bar{\epsilon}_e=$ 6\% and 2.5 \% for $\Delta m^2_{21} = 2 \cdot
10^{-4}$ eV$^2$   
and $10^{-4}$ eV$^2$ respectively. 

Thus, with increase of energy the up-down asymmetry becomes more profound. 
However the absolute value of the excess decreases. 

Let us now consider the multi-GeV  events.  
The zenith angle distributions of the excess for different 
values of $\Delta m^2_{21}$ are shown in Fig.~4.
Let us compare the effects in the multi-GeV and sub-GeV samples:

(i) Typical energy of neutrinos which produce multi - GeV events,  
$E_{mG} \sim  3 - 4$~GeV, is 4 - 5 times larger than  the energy,
$E_{sG}$,  in the sub-GeV sample. Therefore the mixing 
parameter  and the probability  are suppressed by 
factor $(E_{mG}/E_{sG})^2 \sim 16 - 25 $.  

(ii) The ratio of the fluxes, $r$, for the multi-GeV range
is  about  $r \approx 3 $ (for upward bin), 
so that the value of  factor  
$\overline{r}(\Theta_e) c_{23}^2 - 1$ 
turns out to be 2 - 3 times larger than in the sub-GeV sample. 

(iii) Since the average angle between the neutrino and produced charge lepton 
is smaller in the multi-GeV range, the averaging effect is smaller. 
In particular, in the 
vertical bin the relative contribution of trajectories which cross the
core of the Earth is larger. The 
transition probability for the core crossing trajectories is slightly
enhanced. 

(iv) As the result of the interplay of these factors, the excess
in the multi-GeV range is 5 - 7 times smaller  than in the sub-GeV region.

The zenith angle dependence of the excess is stronger. The excess
decreases with $|\cos \Theta|$ and it disappears for the horizontal 
bin. The up-down asymmetry due to oscillations can reach 
$A^{U/D}_{osc} = 2$ \% for $\Delta m^2_{21} = 2\cdot 10^{-4}$ eV$^2$ and 
it decreases rapidly with $\Delta m^2_{21}$. For $\Delta m^2_{21} = 10^{-4}$
eV$^2$ the excess and the asymmetry are below 1 \%.

\section{$\mu$ -like events and the Double Ratio}

In contrast to electron neutrinos, the muon neutrinos 
have both 
small and large $\Delta m^2$ modes of oscillations.   
According to Eq.~(\ref{fluxmu}) the flux 
can be written as 
\be
\frac{F_{\mu}}{F_{\mu}^0} = 
1 -  \sin^2 2\theta_{23} \sin^2 \frac{\phi}{2} - \epsilon_{\mu} - 
\epsilon_{int} ,
\label{fluxmu1}
\ee
where the first two terms correspond to the standard 
$\nu_{\mu} \leftrightarrow \nu_{\tau}$ survival 
probability with slightly modified phase (see below); 
\be 
\epsilon_{\mu} = 
\frac{c_{23}^2}{r} \left({r c_{23}^2} - 1 \right) P_2~ 
\approx \frac{c_{23}^2}{r} \epsilon_e  
\label{epsmu}
\ee
is the correction due to 
$\nu_{\mu} - \nu_e$ oscillations and 
\be 
\epsilon_{int} = 
( 1 - \sqrt{1 - P_2})\sin^2 2\theta_{23} \cos \phi  
\approx \frac{1}{4} P_2 \sin^2 2\theta_{23} \cos \phi~ 
\label{epsint}
\ee
is the term which describes the interference of the oscillations with 
large and small $\Delta m^2$.  
The phase $\phi$ can be estimated in the limit of matter dominance as 
\be 
\phi \approx (\Delta m^2_{31} + s_{12}^2 \Delta m_{21}^2)\frac{L}{2E} . 
\label{phase1}
\ee
The first term here is the standard vacuum oscillation phase, whereas the
second term is the correction due to 3$\nu$ mixing. For the
sub-GeV sample this correction can be neglected. 
Indeed, in  the upward bins 
the oscillations due to $\Delta m^2_{31}$
are averaged and an additional small contribution from 
the second term (associated to $\Delta m_{21}^2$)
play no role. For downward  bins  with the average distance  
$L \sim 100$ km the second term is negligible.   

Due to strong averaging effect the correction 
$\epsilon_{int}$ can be neglected. 
As the result, the number of $\mu$-like events 
can be written as 
\be 
N_{\mu} = N_{\mu}^0 
[\overline{P}(\nu_{\mu}\leftrightarrow \nu_{\tau}) 
- \bar{\epsilon}_{\mu}] ,
\ee
where $\overline{P}$ is the averaged (over energy and zenith angle) 
two neutrino probability. For maximal mixing the 
correction $\bar{\epsilon}_{\mu}$ is very small. 
With decrease of $\sin^2 2\theta_{23}$ the survival probability 
increases $\Delta P \sim 1/2 \Delta \sin^2 2\theta_{23}$. 
At the same time, the correction $\bar{\epsilon}_{\mu}$ increases too,
thus partly compensating of $\Delta{P}$.
The compensation effect depends strongly on $\Delta m^2_{21}$
and $\sin^2 2\theta_{23}$.  
For $\Delta m^2_{21} = 10^{-4}$ eV$^2$  and 
$\sin^2 2\theta_{23} = 0.8$ we find  
$\bar{\epsilon}_{\mu} = 2 $~\%, whereas $\Delta P = 10$~\%. 
For $\sin^2 2\theta_{23} = 0.9$ the corresponding numbers are 
$\bar{\epsilon}_{\mu} = 1.5 $ \% and $\Delta P = 5$ \%.

The  double ratio $R_{\mu/e} \equiv  (N_{\mu}/N_{e})(N_{\mu}^0/N_{e}^0)$ 
can be written as 
\be 
R_{\mu/e} = R_{\mu/e}^{max} 
\frac{1 - 0.5 \sin^2 2\theta_{23} - \bar{\epsilon}_{\mu}}
{1 + \bar{\epsilon_e}}     \;\; , 
\label{doublra}
\ee
where $R_{e/\mu}^{max}$ is the double ratio of two neutrino
oscillations with maximal mixing. In the double ratio both corrections 
$\bar{\epsilon}_{\mu}$  and $\bar{\epsilon_e}$ 
compensate the decrease $\sin^2 2\theta_{23}$. For $\sin^2 2\theta_{23} = 0.9$ 
we find $\Delta P = 5$ \%,
 $\bar{\epsilon}_{\mu} = 1.5 $ \% and 
$\bar{\epsilon}_{e} = 3$ \%,  so that the total increase of the 
double ration is very small: $0.5$\%. 

The corrections $\bar{\epsilon}_{\mu}$ and $\bar{\epsilon}_{e}$ 
are substantially smaller in the multi-GeV 
range,  where one would expect 
$R_{\mu/e}^{mG} >  R_{\mu/e}^{sG}$.

\section{Discussion and Conclusions}

We  have considered the oscillation effects 
in the atmospheric neutrinos induced by 
$\Delta m^2_{21}$ and $\sin^2 2\theta_{21}$ from the region of large mixing
MSW solution of the solar neutrino problem. 
 

The oscillations can lead to the observable 
excess of the e-like events in the sub-GeV sample 
with the following properties. 
The maximal excess is in the low energy part of the sub-GeV sample: 
the integrated excess can range from 
12 \% to 3\% for $\Delta m^2$ decreasing from 
$2 \cdot 10^{-4}$ to $4 \cdot 10^{-5}$ eV$^2$. 
The effect decreases with increase of energy of the 
sample. For high energy part of the 
sub-GeV range the integrated excess can reach 6 \%. 
For the multi-GeV sample the excess is below 1.5 \%.

Notice that the excess of the e-like events can also be due to 
oscillations induced by the large $\Delta m^2_{31}$ responsible for the  
leading channel  of the atmospheric neutrino oscillations, 
provided  that there is some admixture of the $\nu_e$ - flavor in the  
$\nu_3$ state. 
The two effects differ by the energy dependence.   
The excess due to large $\Delta m^2$ oscillations increases with energy:
the excess should be substantially 
stronger in the multi-GeV sample~\cite{ADLS}.
The latest SK data show stronger 
effect in the sub-GeV sample (especially in the upward going bins) thus 
preferring the solar $\Delta m^2$ effect.

The excess due to the solar $\Delta m^2$ oscillations has 
certain   zenith angle dependences: 
The up-down asymmetry due to oscillations is very weak in the low energy
part of the sub-GeV  sample, and it is strong in the high energy 
part of the sample. In the multi-GeV range the asymmetry is also strong, 
however the excess itself is much weaker. 

These properties  will allow one to distinguish the oscillation effect from 
the normalization of fluxes (especially  in future high statistics and 
high precision experiments).

The excess  depends on the mixing angle 
$\sin^2 2\theta_{23}$ responsible for the leading  channel of 
oscillations. The effect decreases with $\sin^2 2\theta_{23}$ 
and it is strongly suppressed for maximal mixing,  
$\sin^2 2\theta_{23} = 1$. Therefore, it is  impossible to exclude the
large mixing  MSW  solution from the
atmospheric neutrino data or even put some bounds on parameters, 
unless the $\sin^2 2\theta_{23}$ will be measured with a
good precision. 
On the contrary, if the excess with  described properties will be
established,   
this will both  confirm  the large mixing MSW solution
and  show that $\sin^2 2\theta_{23}$ differs from 1.  

In this paper we  discussed the excess of the e-like events. 
For $\theta_{23} > 45^{\circ}$ one expects the suppression of the 
e-like events which, in fact, disfavored by the present data. Therefore, 
if $\Delta m_{21}^2$ will be further restricted by the solar
neutrino observations, one  will be able to put the bound on 
$\sin^2 2\theta_{23}$.

Comparison of the predicted excess with data (Fig. 1 - 5)  
shows reasonable  agreement.  
For $\Delta m^2_{21}  \sim 10^{-4}$ eV$^2$ the excess due to oscillations  
reproduces  both the size  and  the zenith angle distribution of the
observed excess in the sub-Gev range.   
It also gives reasonable description of the data in the 
low energy part of the sub-GeV 
sample. At the same time,  the 
predicted excess is smaller  than the detected one  
in the high energy sub-GeV range, and also in the multi-GeV
range. 
Notice, however,  that in the high energy sub-GeV range,  
 and especially  in multi-GeV range, 
the effects of the large $\Delta m^2_{31}$ (neglected here) 
can be important \cite{ADLS}. 
They can change the signal by  15 \% in the  multi-GeV sample 
and by  5 - 6~\% in the high energy part of the sub-GeV sample. 
Also non-trivial interference of small and large 
$\Delta m^2$  oscillation effects is possible~\cite{RS}.   
It is clear however, that  with present   
statistics it is impossible to make 
definite conclusions. 
In fact, it could be that the  explanation of the data 
will  require some  interplay of 
the normalization of fluxes  and  different  oscillation effects. 

The excess has the  positive up-down asymmetry.  
The positive asymmetry at low energies can be reconciled with 
the negative asymmetry at high energies (multi-GeV sample) indicated
by the present data. 
Indeed,  the excess  due to  solar $\Delta m^2$ (with positive 
asymmetry) decreases with increase of energy and 
in the multi-GeV range the dominant effect  
will be due 
to oscillations with large $\Delta m^2$ and   nonzero
$\theta_{13}$.  
In this case the sign of the  excess 
is determined by factor \cite{ADLS}: 
\be
(r s_{23}^2 - 1)
\label{fact}
\ee
(with $s_{23}^2$ instead of $c_{23}^2$, see Eq.~(\ref{fluxe})). 
In the multi-GeV range the average ratio of the fluxes 
$r \approx 2.5$,   and  for $s_{23}^2 = 0.277$ 
we get negative value of the factor (\ref{fact}): $- 0.3$. 
Indeed, the latest SK data indicate a negative 
up-down asymmetry at high energies.  The interpretation of the 
multi-GeV zenith angle distribution can require 
some interplay of the normalization which  explains the excess 
of  events in the down-going bins ($\cos \Theta = 0.2 \div 1$) 
and the oscillations which suppress the number of events in the upward 
going bins ($\cos \Theta = - 1 \div - 0.2$).  

Further studies of solar neutrinos  
(in particular, searches for recoil electron spectrum distortion and
the day-night effect) will allow one to prove or disprove the large
mixing MSW  solution.  Observations  of significant day-night effect
will imply small values of $\Delta m^2_{21}$ and therefore small
excess of the e-like events. On the contrary, weak day-night effect
and  upturn of the distortion of the recoil electron spectrum at low
energies will testify for large $\Delta m^2_{21}$ and large excess.  
The KAMLAND experiment \cite{KAMLAND} will test  whole the range 
of the large mixing angle solution. 

If the effects discussed in this paper will be confirmed we can be left  
with  the ``bi-large  mixing scheme'' with 
large (but not maximal) mixing between neighboring generations and small
$e - \tau$ mixing. The mass hierarchy will  be  rather weak: 
$m_2/m_3 \sim 0.1 - 0.3$ which will allow one to explain large mixing 
without special arrangements.\\ 

{\Large \bf  Note added}

After this work had been accomplished 
the paper~\cite{Giunti} has appeared in which the 
implications of the low-energy SK data 
for LMA solution of the solar neutrino problem are discussed. 
It is claimed that oscillations lead always to decrease of number of
e-like events in comparison with no oscillation case. 
This statement contradicts our results.
The error follows from using the ratio of numbers of events,
$R_{\mu/e}^{MC}$, instead of ratio of the neutrino fluxes ($r$ in our
notation)  in formulas~(2.22)-(3.9) of Ref.~\cite{Giunti}.  Also
matter effects must be included.


\section*{Acknowledgments}

This work was supported by DGICYT under grant PB95-1077 and by the TMR
network grant ERBFMRXCT960090 of the European Union. A. Yu. S.  thanks
Fundacion Bilbao Viscaya (FBBV) for supporting his visit to Valencia
when our collaboration started.
O.L.G. Peres thanks to M.C. Gonzalez-Garcia for
the atmospheric neutrino code and for the extensive joint work made
in  collaboration with H. Nunokawa.



\newpage
\centerline{\large Figure captions}

\vglue 0.4cm
\noindent
\noindent
Fig.~1.  Zenith angle distribution of the excess of the e-like events 
in  the sub-GeV range for $\sin^2 2\theta_{23} = 0.8$, 
$\sin^2 2\theta_{12} = 0.9$ and 
different values $\Delta m^2_{21}$.   
The points are the 735 days data  of
the Super-Kamiokande~\cite{skwin99}.

\noindent
Fig.~2.  Zenith angle distribution of the excess of the e-like events
in  the sub-GeV range for different values of
$\sin^2 2\theta_{12}$ and for two values of 
$\Delta m^2_{21} = 4\; 10^{-5}$~eV$^2 $ and $2\; 10^{-4}$~eV$^2 $;
$\sin^2 2\theta_{23} = 0.8$.
The points are the 735 days data  of
the Super-Kamiokande~\cite{skwin99}.

\noindent
Fig.~3.  The same as in Fig.1 for the low energy part of the sub-GeV range
($p < 0.4$ GeV).

\noindent
Fig.~4.  The same as in Fig.1 for the high energy part of the sub-GeV range
($p > 0.4$ GeV).

\noindent
Fig.~5.  The same as in Fig.1 for the multi-GeV range.

\begin{figure}[ht]
\centering\leavevmode
\epsfxsize=0.8\hsize
\epsfbox{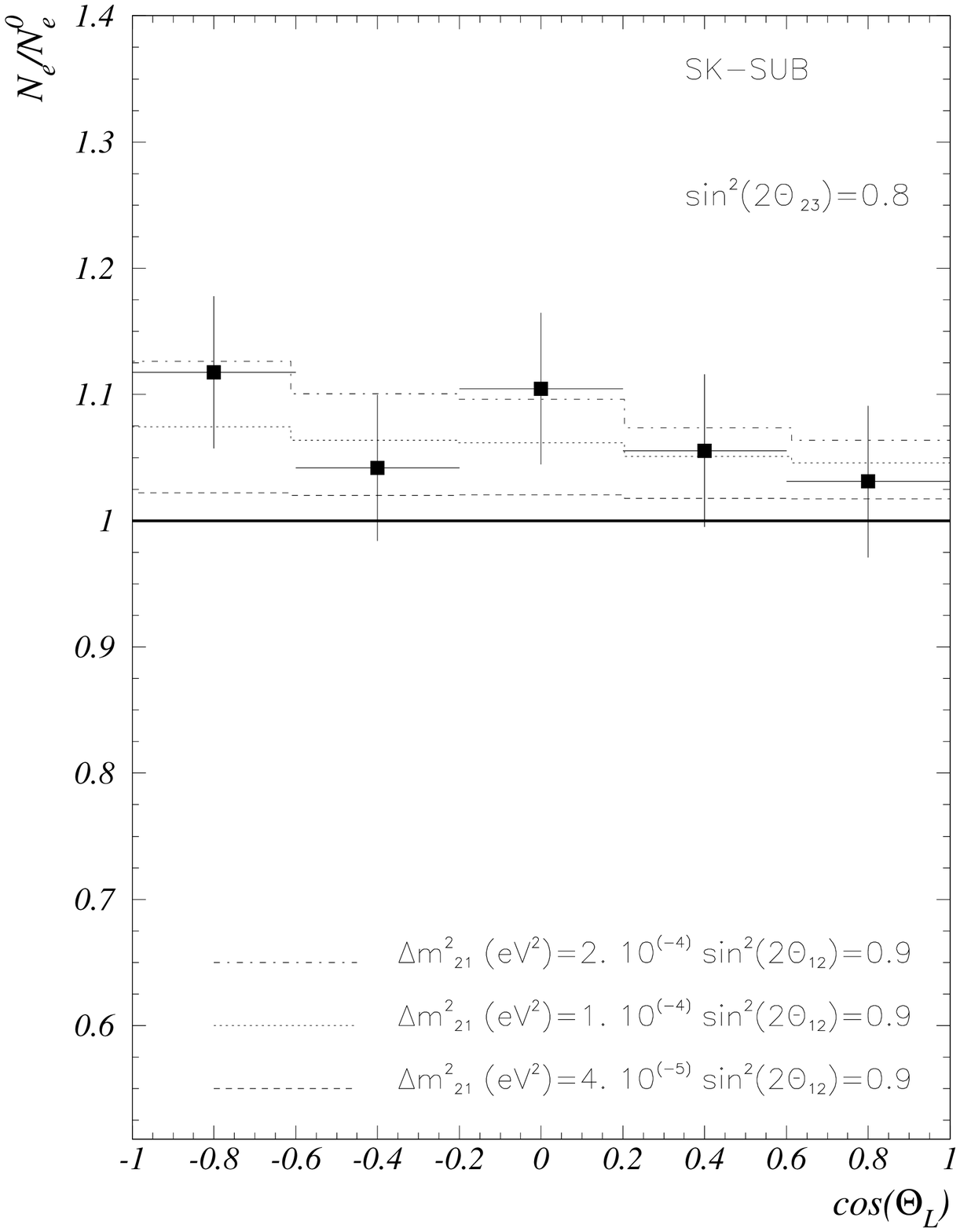}
\caption{
}
\label{fig0} 
\end{figure}
\begin{figure}[ht]
\centering\leavevmode
\epsfxsize=0.8\hsize
\epsfbox{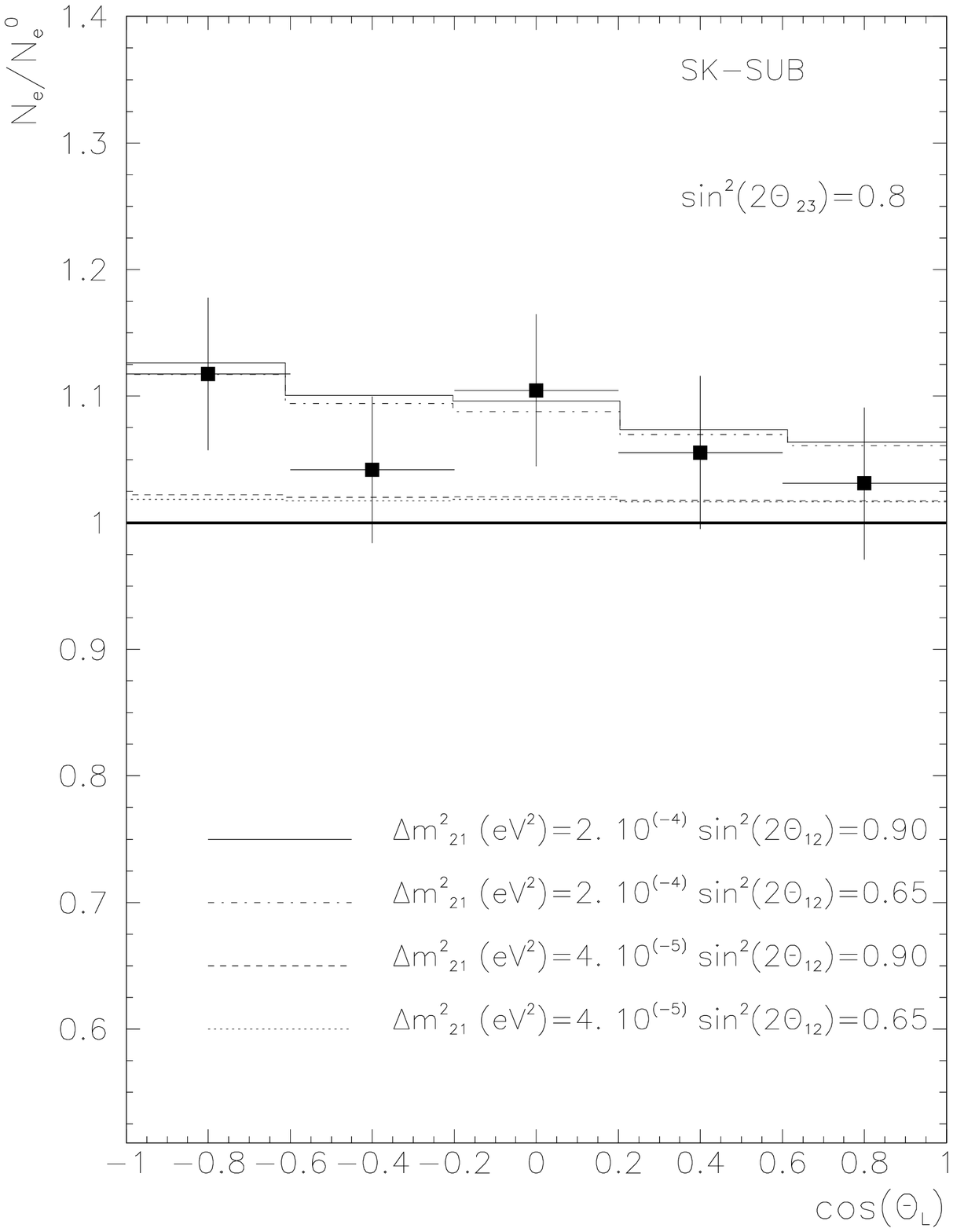}
\caption{
}
\label{fig1} 
\end{figure}

\begin{figure}[ht]
\centering\leavevmode
\epsfxsize=0.8\hsize
\epsfbox{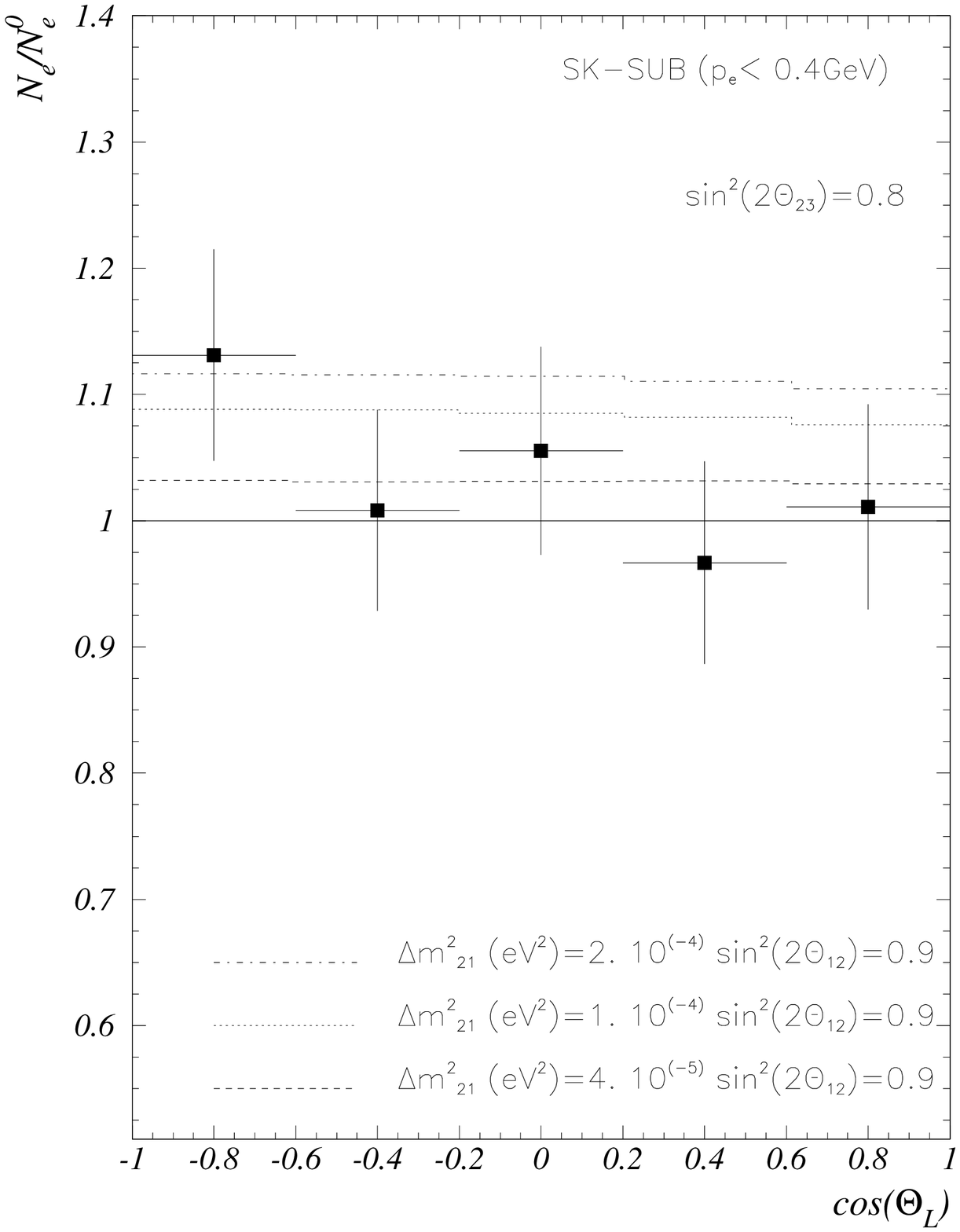}
\caption{
}
\label{fig2} 
\end{figure}

\begin{figure}[ht]
\centering\leavevmode
\epsfxsize=0.8\hsize
\epsfbox{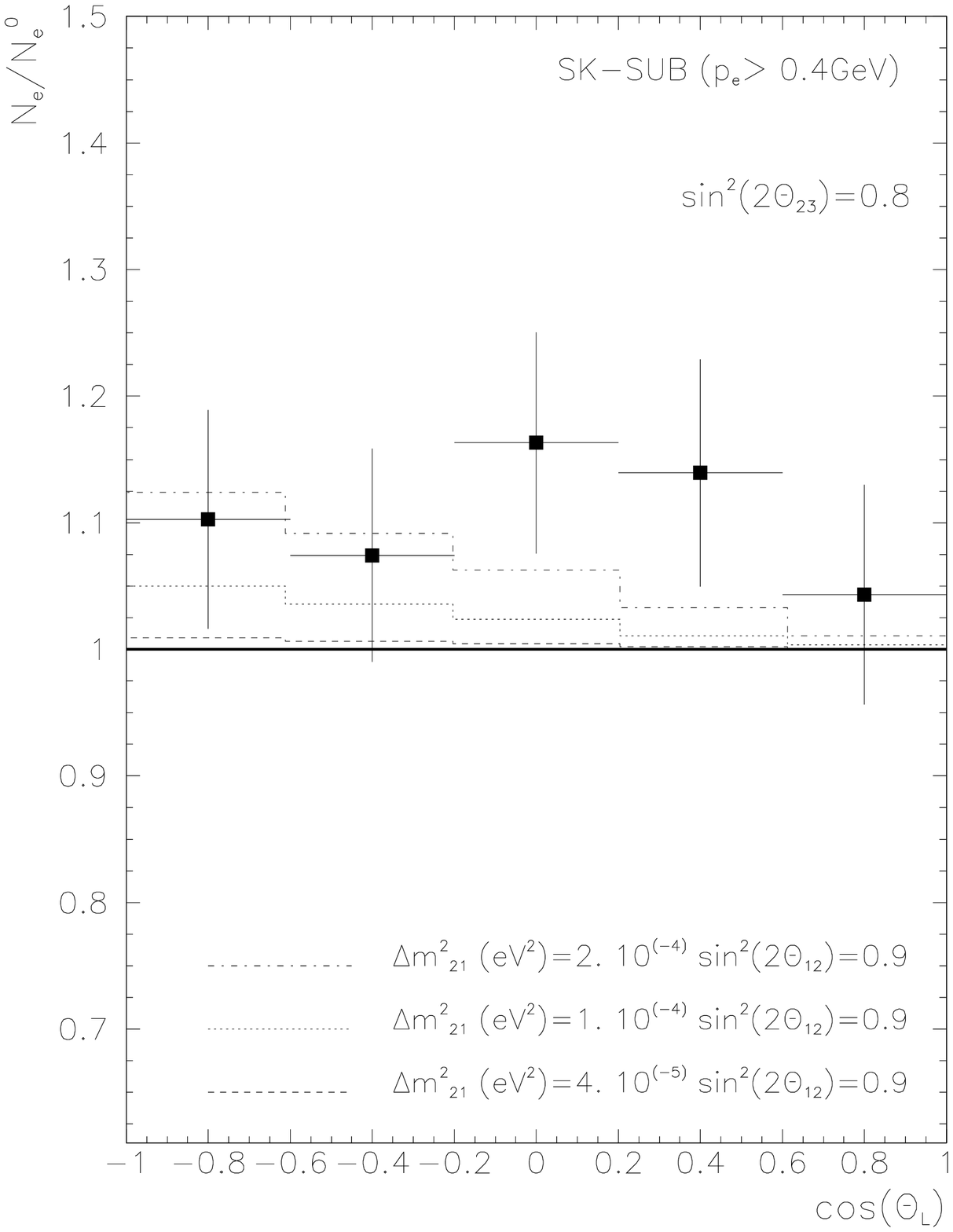}
\caption{
}
\label{fig3}

\end{figure}

\begin{figure}[ht]
\centering\leavevmode
\epsfxsize=0.8\hsize
\epsfbox{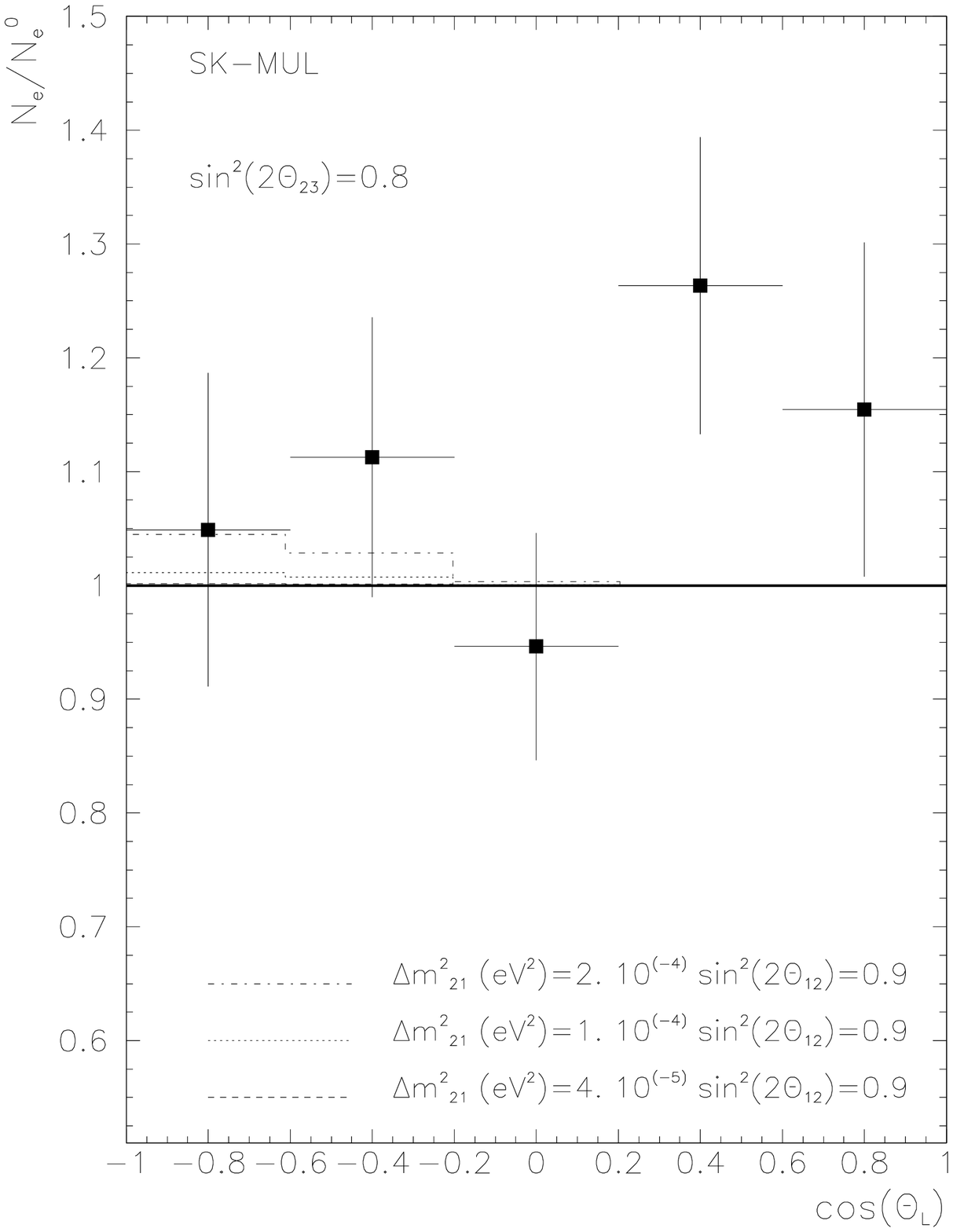}
\caption{
}
\label{fig4}

\end{figure}

\end{document}